\documentclass[final,5p,times,twocolumn]{elsarticle}

\usepackage[T1]{fontenc}
\usepackage[utf8]{inputenc}

\usepackage{amsmath}
\usepackage{amsfonts}
\usepackage{amssymb}
\usepackage{amsxtra}
\usepackage{array}
\usepackage{graphicx}
\usepackage{hepunits}
\usepackage{units}
\usepackage{color}
\usepackage[linkcolor=blue,citecolor=blue,urlcolor=blue,colorlinks=true,breaklinks]{hyperref}
\usepackage{xspace}
\usepackage[normalem]{ulem}

\usepackage[mathscr,scaled=1.15]{urwchancal}

\def\assymto#1{\mbox{\raisebox{-1.2ex}[0.ex][1.6ex]{$\stackrel{\simeq}{\scriptscriptstyle #1}$}}}
\def\eqwhen#1{\mbox{\raisebox{-1.2ex}[0.ex][1.6ex]{$\stackrel{=}{\scriptscriptstyle #1}$}}}

\newcommand{\be}{\begin{equation}}
\newcommand{\bea}{\begin{eqnarray}}
\newcommand{\ee}{\end{equation}}
\newcommand{\eea}{\end{eqnarray}}

\def\1eq#1{Eq.~(\ref{#1})}

\def\2eqs#1#2{Eqs.~(\ref{#1}) and~(\ref{#2})}
\def\3eqs#1#2#3{Eqs.~(\ref{#1}),~(\ref{#2}) and~(\ref{#3})}

\newcommand{\ie}{\textit{i.e. }}

\newcommand{\fatg}{{\rm{I}}\!\Gamma}
\newcommand{\Gnp}{\Gamma}
\newcommand{\overbar}[1]{\mkern 2mu\overline{\mkern-3.5mu #1 \mkern-1.2mu}\mkern 1.2mu}
\newcommand{\blambda}{ \overbar{\lambda} }  

\biboptions{sort&compress}

\journal{Physics Letters B}

\begin{document}

\begin{frontmatter}

\title{Infrared facets of the three-gluon vertex}

\author[UNICAMP]{A.~C.~Aguilar}
\author[UPO]{F.~De Soto}
\author[UNICAMP]{M.~N. Ferreira}
\author[UV]{J.~Papavassiliou}
\author[Huelva]{J.~Rodr\'iguez-Quintero}
\address[UNICAMP]{\mbox{University of Campinas - UNICAMP, Institute of Physics ``Gleb Wataghin,''} 
13083-859 Campinas, S\~{a}o Paulo, Brazil}
\address[UPO]{Dpto. Sistemas F\'isicos, Qu\'imicos y Naturales, Univ. Pablo de Olavide, 41013 Sevilla, Spain}
\address[UV]{Department of Theoretical Physics and IFIC, University of Valencia and CSIC, E-46100, Valencia, Spain}
\address[Huelva]{Dpto. Ciencias Integradas, Centro de Estudios Avanzados en Fis., Mat. y Comp., Fac. Ciencias Experimentales, Universidad de Huelva, Huelva 21071, Spain}

\begin{abstract}
  We  present novel  lattice results  for  the form  factors of  the
  quenched three-gluon vertex of QCD, in two special kinematic configurations
  that  depend on  a single  momentum scale.   We consider  three form
  factors, two  associated with a  classical tensor structure  and one
  without tree-level  counterpart, exhibiting markedly different
  infrared  behaviors.   Specifically,  while the  former  display  the
  typical suppression driven by  a negative logarithmic singularity at
  the  origin, the  latter  saturates at  a  small negative  constant.
  These exceptional  features are analyzed within  the Schwinger-Dyson
  framework,  with the  aid  of special  relations  obtained from  the
  Slavnov-Taylor identities of the theory. The emerging picture of the
  underlying  dynamics  is  thoroughly  corroborated  by  the  lattice
  results, both qualitatively as well as quantitatively.
\end{abstract}


\end{frontmatter}

\section{Introduction}

The  three-gluon  vertex is a central component of QCD\,\mbox{\cite{Marciano:1977su,Ball:1980ax,Davydychev:1996pb}}, being intimately linked to a variety of fundamental nonperturbative phenomena, and the scrutiny of its properties has received considerable attention in recent years\,\cite{Alkofer:2004it,Cucchieri:2006tf,Cucchieri:2008qm,Huber:2012zj,Pelaez:2013cpa,Aguilar:2013vaa,Blum:2014gna,Eichmann:2014xya,Mitter:2014wpa,Williams:2015cvx,Blum:2015lsa,Cyrol:2016tym,Athenodorou:2016oyh,Duarte:2016ieu,Corell:2018yil,Boucaud:2017obn,Aguilar:2019jsj,Aguilar:2019uob,Aguilar:2019kxz,Vujinovic:2018nqc}. 
A particularly noteworthy feature of this vertex in the Landau gauge is the
infrared behavior of the form factors associated with the classical (tree-level) tensorial structures.
Specifically, as the space-like momenta decrease from the ultraviolet to the infrared regime, 
the size of these form factors is gradually reduced, displaying the so-called ``infrared suppression''\,\cite{Cucchieri:2006tf,Cucchieri:2008qm,
Huber:2012zj,Pelaez:2013cpa,Aguilar:2013vaa,Blum:2014gna,Eichmann:2014xya,Mitter:2014wpa,Williams:2015cvx,Blum:2015lsa,Cyrol:2016tym,Athenodorou:2016oyh,Duarte:2016ieu,Corell:2018yil,Boucaud:2017obn,Aguilar:2019jsj,Aguilar:2019uob,Aguilar:2019kxz}. This suppression culminates with the manifestation of 
a logarithmic divergence at the origin, which drives the form factors to negative infinity\,
\cite{Aguilar:2013vaa,Athenodorou:2016oyh,Boucaud:2017obn,Aguilar:2019jsj,Aguilar:2019uob,Aguilar:2019kxz}

As has been explained in detail in the recent literature, this special behavior of the vertex originates from 
the interplay between dynamical effects occurring in the two-point sector of QCD\,\cite{Aguilar:2019jsj,Aguilar:2019uob,Aguilar:2019kxz}.  
In particular, while the gluon acquires dynamically an effective mass\,\cite{Cornwall:1981zr,Bernard:1982my,Donoghue:1983fy,Wilson:1994fk,Philipsen:2001ip}, responsible for the infrared saturation of the Landau-gauge gluon propagator\,\cite{Cucchieri:2007md,Bogolubsky:2007ud,Bogolubsky:2009dc,Oliveira:2009eh,Ayala:2012pb,Aguilar:2004sw,Aguilar:2006gr,Aguilar:2008xm,Boucaud:2008ky,Fischer:2008uz,Dudal:2008sp,RodriguezQuintero:2010wy,Tissier:2010ts,Pennington:2011xs,Cloet:2013jya,Fister:2013bh,Cyrol:2014kca,Binosi:2014aea,Cyrol:2018xeq}, the ghost remains massless even nonperturbatively\,\cite{Alkofer:2000wg,Fischer:2006ub,Aguilar:2008xm,Boucaud:2008ky,Boucaud:2008ji}.
As a result, loop diagrams containing ghost propagators furnish infrared divergent logarithms, while
gluonic loops, being ``protected'' by the mass, are infrared finite.

The formalism
obtained from the fusion of the Pinch Technique~\cite{Cornwall:1981zr,Cornwall:1989gv,Pilaftsis:1996fh,Binosi:2009qm} with the Background Field Method (PT-BFM)~\cite{Abbott:1980hw},
known as “PT-BFM'' scheme~\cite{Binosi:2009qm,Aguilar:2006gr,Binosi:2007pi}, is particularly suitable for 
exposing this interplay, by combining the
Schwinger-Dyson equations (SDEs) with the
Slavnov-Taylor identity (STI) satisfied by the three-gluon vertex\,\cite{Marciano:1977su,Ball:1980ax,Davydychev:1996pb,Gracey:2019mix}.

In the present work we employ this scheme to 
scrutinize further this dynamical picture, through the analysis of new results
from quenched lattice simulations for three vertex form factors, defined in  
two special kinematic configurations that involve a single momentum scale.

In particular, in the case of the two ``classical'' form factors simulated, 
a considerable increase in the statistics permits us to obtain a cleaner signal 
of the infrared divergences that they display, and accurately determine their strength.
This new information, in turn, enables us to probe more stringently, at the quantitative level,
the underlying mechanisms associated with their emergence.

In addition, we present for the first time lattice results for a form
factor that has no classical analogue. The dynamics of this purely quantum contribution may be described by means of the corresponding 
SDE, and, in contradistinction to the classical form factors, 
does not display any infrared divergences. The data obtained corroborate this prediction, being completely
compatible with a finite rather than a divergent contribution at low momenta.

\section{General considerations and theoretical setup}

Our point of departure is the three-point correlation function composed by SU(3) gauge fields, $\widetilde{A}^a_\alpha(q)$,
in Fourier space, 
\be
{\cal G}_{\alpha \mu \nu}^{abc}(q,r,p) \ = \ \langle \widetilde{A}^a_\alpha(q) \widetilde{A}^b_\mu(r) \widetilde{A}^c_\nu(p) \rangle \ = \ 
f^{abc} {\cal G}_{\alpha \mu \nu}(q,r,p) \, , \label{eq:Green3g}
\ee 
with $q+r+p=0$.\, ${\cal G}_{\alpha \mu \nu}^{abc}(q,r,p)$ may be cast in the form 
\be\label{eq:gGD3}
{\cal G}_{\alpha \mu \nu}(q,r,p) = g \overline{\Gamma}_{\alpha \mu \nu}(q,r,p) \Delta(q^2) \Delta(r^2) \Delta(p^2)  \ , 
\ee
where we have introduced the {\it transversally projected vertex}\,\cite{Aguilar:2019uob,Aguilar:2019kxz},
\be\label{eq:Gammabar}
\overline{\Gamma}_{\alpha \mu \nu}(q,r,p) \ = \  \fatg^{\alpha' \mu' \nu'}(q,r,p) P_{\alpha' \alpha}(q) P_{\mu' \mu}(r) P_{\nu' \nu}(p)  \ ,
\ee
with $\fatg$ denoting the usual one-particle irreducible (1PI) three-gluon vertex.
In addition, $g$ is the gauge coupling, and $\Delta(q^2)$ the scalar component of the gluon propagator,  
\be\label{eq:Delta}
\Delta^{ab}_{\mu\nu}(p) = \langle \widetilde{A}^a_\mu(p) \widetilde{A}^b_\mu(-p) \rangle = \Delta(p^2) \delta^{ab} P_{\mu\nu}(p) \,,
\ee
with $P_{\mu\nu}(p)=g_{\mu\nu} - p_\mu p_\nu/p^2$, the standard transverse projector. 
Evidently, $q^\alpha {\cal G}_{\alpha\mu\nu} = r^\mu {\cal G}_{\alpha\mu\nu} = p^\nu {\cal G}_{\alpha\mu\nu}=0$.

The typical quantity studied in Landau-gauge lattice simulations is the projection\,\cite{Athenodorou:2016oyh,Duarte:2016ieu,Boucaud:2017obn}
\be
L(q,p,r;\lambda) =  \frac{{\cal G}_{\alpha \mu \nu}(q,r,p) \lambda^{\alpha \mu \nu}(q,r,p)}
{\rule[0cm]{0cm}{0.45cm}\lambda_{\alpha \mu \nu}(q,r,p) \lambda^{\alpha \mu \nu}(q,r,p)} 
\rule[0cm]{0cm}{0.5cm} \;,
\label{eq:latobs}
\ee
where $\lambda^{\alpha\mu\nu}$ is a transverse tensor whose form should be appropriately chosen, depending on the kinematic configuration employed 
and  the form factor that one wants to extract.
In what follows we will focus our
attention on the {\it (i)} \emph{totally symmetric} and {\it (ii)} \emph{asymmetric} configurations of the three-gluon vertex. 

In case {\it (i)}, the momenta configuration is defined by \mbox{$q^2=p^2=r^2:=s^2$}, such that \mbox{$q\cdot r = q\cdot r = r \cdot p = - s^2/2$}
and \mbox{$\theta = \widehat{q r} = \widehat{q p} = \widehat{r p} = 2\pi/3$.} The tensor structure of $\overline{\Gamma}$ is then reduced down to\,\cite{Alles:1996ka,Boucaud:1998bq}
\be\label{eq:Gamma1and2}
\overline{\Gamma}^{\alpha \mu \nu}(q,r,p) =  \overline{\Gamma}^{\rm sym}_1(s^2) \, \lambda_1^{\alpha \mu \nu}(q,r,p) + 
\overline{\Gamma}^{\rm sym}_2(s^2) \, \lambda_2^{\alpha \mu \nu}(q,r,p)  \,, 
\ee
with the two tensors
\begin{subequations}
\label{eq:symtensors}
\begin{align}
\label{eq:lambda1}
\lambda_1^{\alpha \mu \nu}(q,r,p) &= \overline{\Gamma}_0^{\alpha \mu \nu}(q,r,p)\,,  \\
\lambda_2^{\alpha \mu \nu}(q,r,p) &= \frac{(q-r)^\nu (r-p)^\alpha (p-q)^\mu}{s^2} \ ;
\label{eq:lambda2}
\end{align}
\end{subequations}
$\overline{\Gamma}_0^{\alpha \mu \nu}(q,r,p)$ is the tree-level version of the vertex in Eq.\,(\ref{eq:Gammabar}).

We next project out of $L$ two particular combinations, denoted by $T_i^{\rm sym}$,
each containing one of the $\overline{\Gamma}_i^{\rm sym}$, namely
\be\label{eq:gGsym}
T^{\rm sym}_i(s^2) := g \overline{\Gamma}^{\rm sym}_i(s^2) \Delta^3(s^2) =
\left.
L\left(\blambda_i\right) \rule[0cm]{0cm}{0.3cm} \right|_{q^2=r^2=p^2:=s^2} \; ,
\ee
where
\be\label{eq:lambdatilde}
\blambda_i^{\alpha \mu \nu}(q,r,p) \ = \ \sum_{j=1}^2 \beta_{ij} \, \lambda_j^{\alpha \mu \nu}(q,r,p) \; ,
\ee
with 
$\beta_{11}=1$, \,$\beta_{12}=1/2$, \, $\beta_{21}=6/11$, and $\beta_{22}=1$, such that
\be
\blambda_{i\,\alpha \mu \nu}(q,r,p) \,\lambda_j^{\alpha \mu \nu}(q,r,p) \ = \ \delta_{ij} \, \blambda_{i\,\alpha \mu \nu}(q,r,p) \, \blambda_i^{\alpha \mu \nu}(q,r,p) \ .
\ee

 In case {\it (ii)}, the asymmetric configuration corresponds to the kinematic limit
\mbox{$p \to 0$}, \mbox{$r = -q$} and  \mbox{$\theta = \widehat{qr}=\pi$}.  In these kinematics we have\,\cite{Boucaud:1998xi} 
\be\label{eq:Gamma3}
\overline{\Gamma}^{\alpha\mu\nu}(q,-q,0) = \overline{\Gamma}_3^{\rm asym}(q^2) \lambda_3^{\alpha\mu\nu}(q,-q,0)  \;,
\ee
in terms of the single tensor, 
\be\label{eq:lambda3}
\lambda_3^{\alpha\mu\nu}(q,-q,0) = 2 q^\nu P^{\alpha\mu}(q) \,,
\ee
which emerges after the implementation of the asymmetric limit on the tensorial basis of the
three-gluon vertex. 

A careful analysis reveals that the projection of  $\overline{\Gamma}^{\rm asym}_3$
from $L$ proceeds through contraction by 
$\lambda_3^{\alpha\mu\nu}(q,-q,0)$ itself, namely 
\be\label{eq:gGasym}
T^{\rm asym}_3(q^2) = g \overline{\Gamma}^{\rm asym}_3(q^2) \Delta^2(q^2) \Delta(0) =
\left. L\left(\lambda_3 \right)
\rule[0cm]{0cm}{0.3cm} \right|_{r^2=q^2;p^2\to 0} \;.
\ee
Note that the limit $p\to 0$ is {\it path-independent}, {\it i.e.}, 
does {\it not} depend on the angle formed between $p$ and $q$.

We next implement multiplicative renormalization by introducing 
the standard renormalization constants, $Z_i$, relating 
bare and renormalized quantities as
\begin{align}
\Delta_R(q^2) &= Z_A^{-1} \Delta(q^2),  &{\cal G}_R(q,r,p) &= Z_A^{-3/2} {\cal G}(q,r,p),  \nonumber 
\\ 
g_R &=  Z_A^{3/2} Z_3^{-1} g, &\overline{\Gamma}_R(q,r,p) &= Z_3 \overline{\Gamma}(q,r,p)\,. 
\label{eq:genren}
\end{align}
Within the momentum subtraction (MOM) scheme\,\cite{Hasenfratz:1980kn} that we use, 
the renormalized correlation functions must acquire
their tree-level expressions at the subtraction point $\mu^2$, {\it e.g.},  \mbox{$\Delta^{-1}_{{\rm R}}(\mu^2) = \mu^2$}.

Turning to the kinematic configurations {\it (i)} and {\it (ii)}, we impose, correspondingly, the MOM conditions
\be\label{eq:renG1}
\overline{\Gamma}_{1\,{\rm R}}^{\rm sym}(\mu^2)  \ = \ 1 \,, \qquad
\overline{\Gamma}_{3\,{\rm R}}^{\rm asym}(\mu^2) \ = \ 1 \,, 
\ee
which define the {\it symmetric} and {\it asymmetric} MOM schemes, respectively\,\cite{Alles:1996ka,Davydychev:1997vh,Boucaud:1998bq,Boucaud:1998xi}.

Focusing on case {\it (i)}, we want to express $\overline{\Gamma}_{{\rm R}\,i}^{\rm sym}(s^2)$
exclusively in terms of the bare lattice quantities $\Delta$ and $T^{\rm sym}_i$.
This may be readily accomplished, since multiplicative renormalization entails that, 
for any correlation function $G(q^2)$, the ratio $G(q^2_1)/G(q^2_2) = G_R(q^2_1)/G_R(q^2_2)$
is a renormalization-group invariant combination. 

In particular,  
forming the ratio $T^{\rm sym}_{i}(s^2)/ T_{1}^{\rm sym}(\mu^2)$ using Eq.\,(\ref{eq:gGsym}), and employing the condition
of Eq.\,(\ref{eq:renG1}), we find 
\be\label{eq:Gfin}
\overline{\Gamma}_{i\,{\rm R}}^{\rm sym}(s^2) = 
\frac{T^{\rm sym}_i(s^2)}{T^{\rm sym}_1(\mu^2)} \, \left( \frac{\Delta(\mu^2)}{\Delta(s^2)} \right)^3  \quad \mbox{with}\quad i=1,2\,.
\ee
Applying exactly analogous reasoning to the case  {\it (ii)}, we obtain 
\be\label{eq:Gasfin}
\overline{\Gamma}_{3\,{\rm R}}^{\rm asym}(q^2) = 
\frac{T^{\rm asym}_3(q^2)}{T^{\rm asym}_3(\mu^2)} \, \left( \frac{\Delta(\mu^2)}{\Delta(q^2)} \right)^2 \,.
\ee
From this point on, we drop the subscript ``${\rm R}$'' from the renormalized $\overline{\Gamma}_{i}$. 

\section{Connecting the two- and three-point sectors of QCD}

In this section we present the salient features of PT-BFM  approach, pertinent to the the gluon propagator and 
three gluon vertex. The upshot of these considerations is the derivation of theoretical expressions for
the form factors $\overline{\Gamma}_{i}$, which will be contrasted with the
new lattice results in the next section. 

Within the PT-BFM framework 
it is natural to cast the infrared finite $\Delta(q^2)$ as the sum of two  distinct pieces~\cite{Binosi:2012sj} ,
\be
\label{eq:gluon_m_J}
\Delta^{-1}(q^2) = q^2J(q^2) + m^2(q^2)\,,
\ee
where $J(q^2)$ denotes the so-called ``kinetic term'', 
while $m^2(q^2)$ represents a momentum-dependent mass scale. Clearly, 
\mbox{$m^2(0) = \Delta^{-1}(0)$} is the saturation point of the gluon propagator.

The emergence of $m^2(q^2)$ hinges crucially on the structure of $\fatg_{\alpha\mu\nu}$, entering  
in the SDE for $\Delta(q^2)$.  
In particular, $\fatg_{\alpha\mu\nu}$ must be decomposed as 
\be
\fatg_{\alpha\mu\nu}(q,r,p)=\Gnp_{\alpha\mu\nu}(q,r,p) + V_{\alpha\mu\nu}(q,r,p)\,,
\label{3gdec}
\ee
where $V_{\alpha\mu\nu}$ is comprised by {\it longitudinally coupled massless poles}, 
\ie \mbox{$P^\alpha_{\alpha'}(q)P^\mu_{\mu'}(r)P^\nu_{\nu'}(p)V_{\alpha\mu\nu}(q,r,p) = 0$}, 
and $\Gnp_{\alpha\mu\nu}$ denotes the pole-free part of the vertex.
By virtue of the above property, 
$V_{\alpha\mu\nu}$ drops out from the $\overline{\Gamma}_{\alpha \mu \nu}$ in Eq.\,(\ref{eq:Gammabar}), and, 
consequently, the lattice projection of Eq.\,(\ref{eq:latobs}) depends only on $\Gnp_{\alpha\mu\nu}$.   

$\Gamma^{\alpha \mu \nu}(q,r,p)$ is usually decomposed into a longitudinal and a transverse contribution~\cite{Ball:1980ax,Davydychev:1996pb,Aguilar:2019jsj,Aguilar:2019kxz}
\be
\Gamma^{\alpha \mu \nu}(q,r,p) \ = \ \Gamma_L^{\alpha \mu \nu}(q,r,p) + \Gamma_T^{\alpha \mu \nu}(q,r,p) \;,
\ee
with \mbox{$q_\alpha\Gamma_T^{\alpha \mu \nu}(q,r,p) = r_\mu\Gamma_T^{\alpha \mu \nu}(q,r,p) = p_\nu\Gamma_T^{\alpha \mu \nu}(q,r,p) = 0$}.

Their tensorial decomposition in the basis of~\cite{Ball:1980ax,Davydychev:1996pb,Aguilar:2019jsj},  reads    
\begin{subequations}
\label{eq:GammaLyT}
\begin{align}
\label{eq:GammaL}
\Gamma_L^{\alpha \mu \nu}(q,r,p) &= \sum_{i=1}^{10} X_i(q^2,r^2,p^2) \,\ell_i^{\alpha \mu \nu}(q,r,p) \;, \\
\Gamma_T^{\alpha \mu \nu}(q,r,p) &= \ \sum_{i=1}^{4} Y_i(q^2,r^2,p^2)\, t_i^{\alpha \mu \nu}(q,r,p) \;, 
\label{eq:GammaT}
\end{align}
\end{subequations}
where the explicit form of the basis tensors $\ell_i$ and $t_i$ is given in Eqs.~(3.4) and~(3.5) of~\cite{Aguilar:2019jsj}.

Note that the tree-level expression for $\Gamma^{\alpha \mu \nu}$ is recovered from Eq.\,(\ref{eq:GammaL}) by setting $X_1=X_4=X_7 = 1$, and
$X_i$ = 0 for all remaining terms.

Using the above decomposition, one may express the $\overline{\Gamma}_i$ in terms of the $X_i$ and $Y_i$.
Specifically, in Euclidean space, we obtain  
\bea
\overline{\Gamma}_1^{\rm sym}(s^2) &=&  X_1(s^2) - \frac{s^2} 2 X_3(s^2) + \frac{s^4} 4 Y_1(s^2) - \frac{s^2} 2 Y_4(s^2)  \,,
\nonumber \\ 
\overline{\Gamma}_2^{\rm sym}(s^2) &=&  \frac{3 s^2}{4} X_3(s^2) - \frac{3 s^4}{8} Y_1(s^2) - \frac{s^2}{4} Y_4(s^2) \,,
\label{eq:GsXsYs}
\eea
where $X_i(s^2) \equiv X_i(s^2,s^2,s^2)$ and $Y_i(s^2)\equiv Y_i(s^2,s^2,s^2)$.  Moreover,   one has
\be\label{eq:G3XsYs}
\overline{\Gamma}^{\rm asym}_3(q^2) = X_1(q^2,q^2,0) - q^2 X_3(q^2,q^2,0) \; .
\ee
%

Past this point, we will determine the $X_i$ by 
resorting to a construction relying on the STIs satisfied by $\fatg_{\alpha \mu \nu}$, \ie,
\be
p^\nu \fatg_{\alpha \mu \nu}(q,r,p) = F(p^2) [ {\cal T}_{\mu\alpha}(r,p,q) - {\cal T}_{\alpha\mu}(q,p,r) ] \,,
\label{stip}
\ee
with 
\be
{\cal T}_{\mu\alpha}(r,p,q) := \Delta^{-1}(r^2) P_\mu^\sigma(r) H_{\sigma\alpha}(r,p,q)\,.
\label{defAT}
\ee
$F(p^2)$ denotes the ghost dressing function, while 
$H_{\nu\mu}(q,p,r)$ is the ghost-gluon scattering kernel~\cite{Ball:1980ax,Davydychev:1996pb,Aguilar:2018csq},
whose tensorial decomposition is given by [$A_i\equiv A_i(q,p,r)$]
\be 
H_{\nu\mu}(q,p,r) = g_{\nu\mu} A_1 + q_\mu q_\nu A_2 + r_\mu r_\nu A_3 + q_\mu r_\nu A_4 + r_\mu q_\nu A_5 \,.
\label{theAi}
\ee

The decompositions given in Eqs.\,(\ref{eq:gluon_m_J}) and (\ref{3gdec}) prompt the
separation of the above STI into two ``partial'' STIs, obtained by implementing the matching 
$\Gamma \leftrightarrow J$ and $V \leftrightarrow m^2$~\cite{Aguilar:2011xe,Binosi:2012sj}. Based on this hypothesis,
one may extend the BC construction of~\cite{Ball:1980ax}   
to the case of infrared finite gluon propagator, expressing the $X_i$ in terms of the $J$, the $F$, and the $A_i$.
In particular, we obtain
\begin{align}
X_1(s^2) &= Z_1^{\rm sym} F(s^2)J(s^2)R_1^{\rm sym}(s^2) \,, \nonumber \\
X_3(s^2) &= Z_1^{\rm sym} F(s^2)\left[ J^\prime(s^2)R_2^{\rm sym}(s^2) + J(s^2)R_3^{\rm sym}(s^2) \right] \,, \label{eq:Xisym}
\end{align}
and 
\begin{align}
X_1(q^2,q^2,0) &= Z_1^{\rm asym} F(q^2)J(q^2)R_1^{\rm asym}(q^2) \,, \label{eq:Xiasym} \\
X_3(q^2,q^2,0) &= Z_1^{\rm asym} F(0)\left[ J^\prime(q^2)R_2^{\rm asym}(q^2) + J(q^2)R_3^{\rm asym}(q^2) \right] \,, \nonumber
\end{align}
where the $R_j^{\rm sym}$ and $R_j^{\rm asym}$ are linear combinations of the $A_i$ and their derivatives,
whereas $Z_1^{\rm sym}$ and $Z_1^{\rm asym}$ are, respectively, the \emph{finite} renormalization constants~\cite{Aguilar:2020yni} of the ghost-gluon kernel in the symmetric and asymmetric MOM schemes, defined in Eq.\,\eqref{eq:renG1}.

Note that this procedure leaves the transverse vertex form factors $Y_i$ undetermined; nonetheless, as we will see in the next section,
their qualitative features may be deduced from the corresponding SDE governing the vertex $\Gamma$.  

\begin{figure}[!t]
\includegraphics[width=1.\linewidth]{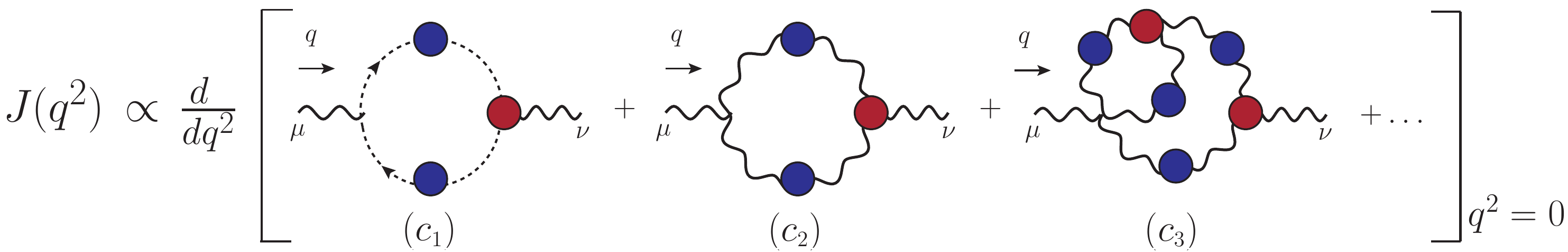}
\vspace*{-0.55cm}
\caption{SDE diagrams contributing to the derivative of the gluon propagator at the origin. Blue (red) circles indicate
fully dressed propagators (vertices).}
\label{fig:3g_sde}
\end{figure}

The ingredients comprising \2eqs{eq:Xisym}{eq:Xiasym} are obtained as follows.
$R_j^{\rm sym}$ and $R_j^{\rm asym}$ may be computed using the SDE results for the form factors $A_i$
presented in~\cite{Aguilar:2019uob}. The values of $Z_1^{\rm sym}$ and $Z_1^{\rm asym}$ have been estimated
by means of a one-loop calculation in~\cite{Aguilar:2020yni}, while $F(q^2)$ is accurately known both from
lattice simulations and functional studies. Finally, the gluon kinetic term $J(q^2)$ requires a more elaborate
treatment, which is outlined below.

\begin{figure}[!th]
\includegraphics[width=1.\linewidth]{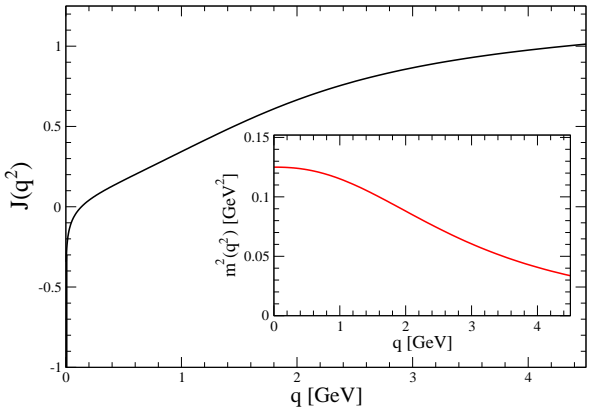}
\vspace*{-0.55cm}
\caption{ Gluon kinetic term, $J(q^2)$, and gluon mass, $m^2(q^2)$ (inset). When combined according to
\1eq{eq:gluon_m_J}, they reproduce accurately the lattice data of~\cite{Bogolubsky:2007ud} for $\Delta(q^2)$.}
\label{fig:J_and_m}
\end{figure}

To determine $J(q^2)$, we  first compute  $m^2(q^2)$ from its own dynamical equation (see, {\it e.g.},~\cite{Aguilar:2019kxz});
the result is shown in the inset of Fig.\,\ref{fig:J_and_m}.
Then, we subtract the $m^2(q^2)$ from the
lattice data for the gluon propagator~\cite{Bogolubsky:2007ud}, by  
employing \1eq{eq:gluon_m_J}, {\it i.e.}, \mbox{$J(q^2) = [\Delta_{{\rm latt}}^{-1}(q^2) - m^2(q^2)]/q^2$}.  
While this procedure is completely stable for a wide range of momenta, 
it becomes less reliable as $q^2\to 0$, due the fact that $J(q^2)$ diverges logarithmically at the origin, {\it e.g.},
\be
J(q^2) \; \assymto{q^2 \to 0} \; a \ln(q^2/\mu^2) + b \,,
\label{Jasympt}
\ee
as a direct consequence of the the nonperturbative masslessness of the ghost~\cite{Aguilar:2013vaa}.

It turns out that the behavior of $J(q^2)$ near the origin may be computed from the SDE of the gluon propagator,
by recognizing that, in the limit $q^2 \to 0$, differentiation with respect to $q^2$ singles out the divergent contribution of $J(q^2)$, {\it e.g.},  
\be 
d\Delta^{-1}(q^2)/d q^2 \eqwhen{q^2 \to 0}  J(q^2) + \ldots \,,
\label{PropDer}
\ee
where the ellipses denote infrared finite terms.

The direct differentiation of the diagrams contributing to 
the SDE of $\Delta(q^2)$ [see Fig.\,\ref{fig:3g_sde}] leads to major technical simplifications,
yielding finally the value of $a \approx 0.046$.
It should be noted that all diagrams $c_i$ contribute to the value of $a$. Specifically, diagram $(c_1)$
furnishes the primary diverge, owing to the masslessness of the ghost propagators,
while $(c_2)$ and $(c_3)$ contribute secondary divergences, due to fully-dressed three-gluon vertices 
attached to the their external leg (Lorentz index $\nu$). 

Thus, the combined treatment furnishes $J(q^2)$ for the entire range of $q^2$, as shown in Fig.\,\ref{fig:J_and_m}. 

\begin{figure}[!th]
\includegraphics[width=1.\linewidth]{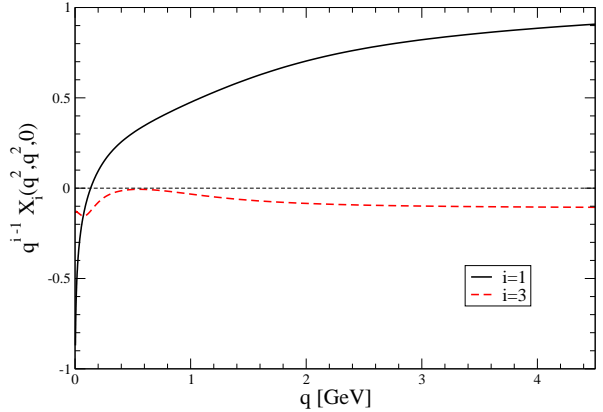} 
\vspace*{-0.55cm}
\caption{ The form factor $X_1(q^2,q^2,0)$ (black continuous) which composes the tree-level tensor structure of the full three-gluon vertex and the dimensionless combination $q^2 X_3(q^2,q^2,0)$ (red dashed).}
\label{fig:X1_X3}
\end{figure}

Finally, putting together all ingredients described above, we obtain from \2eqs{eq:Xisym}{eq:Xiasym} 
the SDE-derived result for the $X_i$ in the two
kinematic configurations of interest. 
In particular, in the asymmetric limit we obtain the $X_1(q^2,q^2,0)$ and $q^2 X_3(q^2,q^2,0)$ shown in Fig.\,\ref{fig:X1_X3}.

As will become apparent in the next section, $X_1$ diverges logarithmically, inheriting directly
from Eqs.\,(\ref{eq:Xisym}) and (\ref{eq:Xiasym}) the corresponding logarithmic divergence of the $J(q^2)$, given by Eq.\,(\ref{Jasympt}).
Instead, while $X_3$ is dominated by the divergent $J^\prime(q^2)$, the combinations $s^2X_3(s^2)$ and $q^2X_3(q^2,q^2,0)$ appearing in Eqs.\,\eqref{eq:GsXsYs} and\,\eqref{eq:G3XsYs} saturate to finite constants in the infrared.

\section{Presentation and analysis of the results}

The lattice evaluation of the form factors 
$\overline{\Gamma}_{1,2}^{\rm sym}$ and $\overline{\Gamma}_{3}^{\rm asym}$
proceeds through 
the direct simulation of the projections $T_i$  and of the gluon propagator $\Delta$ [\3eqs{eq:gGsym}{eq:gGasym}{eq:Delta}],  
and subsequent use of  Eqs.\,\eqref{eq:Gfin} and \eqref{eq:Gasfin}, respectively.
This is accomplished by exploiting lattice gauge field configurations obtained from simulations with the Wilson action on a 48$^4$ lattice at $\beta$=5.8 (970 configurations) and 5.6 (980 configurations), and on a $52^4$ lattice at $\beta$=5.6 (980 configurations);
for further details see~\cite{Boucaud:2018xup}. 
In addition, we have reanalyzed 1050 gauge-field configurations produced with the tree-level
Symanzik action in a $64^4$ lattice~\cite{Athenodorou:2016oyh,Boucaud:2017obn},
making thereby apparent that different discretizations of the QCD action provide practically the same results for the three-gluon form factors. In the case of $\overline{\Gamma}_{1}^{\rm sym}$, for the sake of both comparison and implementation of the renormalization condition
\footnote{For Eq.\,\eqref{eq:Gfin} to work properly , the bare quantities evaluated both at $s^2$ and $\mu^2$ must be computed from configurations simulated at the same $\beta$, such that the cut-off dependence, assumed to be multiplicative, cancels out in the ratios. Therefore, as the accessible momenta to \mbox{$\beta$=5.8 and 5.6} do not reach \mbox{4.3 GeV}, one needs to first fix a renormalization condition at a lower momentum and next match the results to previous data renormalized at 4.3 GeV. As it is obvious from Eq.\,\eqref{eq:Gfin}, the overall matching constant required for $\overline{\Gamma}_1^{\rm sym}$ also applies for $\overline{\Gamma}_2^{\rm sym}$.}
at \mbox{$\mu$=4.3 GeV}, we have also used earlier results~\cite{Boucaud:2017obn}, which cover a wider range of momenta.
All these configurations have been obtained from large-volume, quenched lattice simulations,  neglecting thus the effect of dynamical quarks.
The implications of this approximation has been recently assessed in~\cite{Aguilar:2019uob}, where 
only minor quantitative but no qualitative effects have been detected. 

\begin{figure}[!th]
\begin{tabular}{c}
\includegraphics[width=1.05\linewidth]{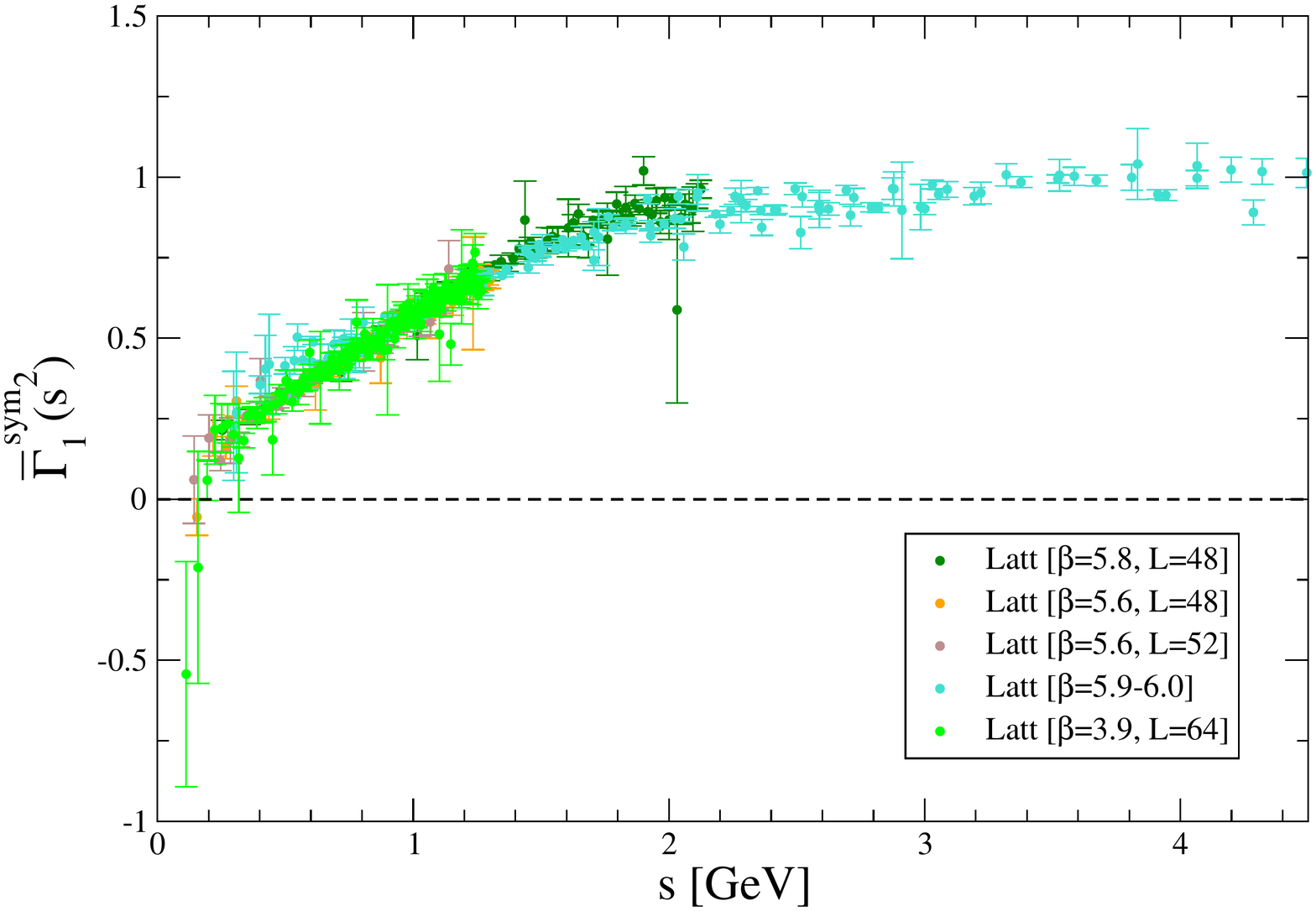} 
\vspace*{-1cm}
\\
\includegraphics[width=1.05\linewidth]{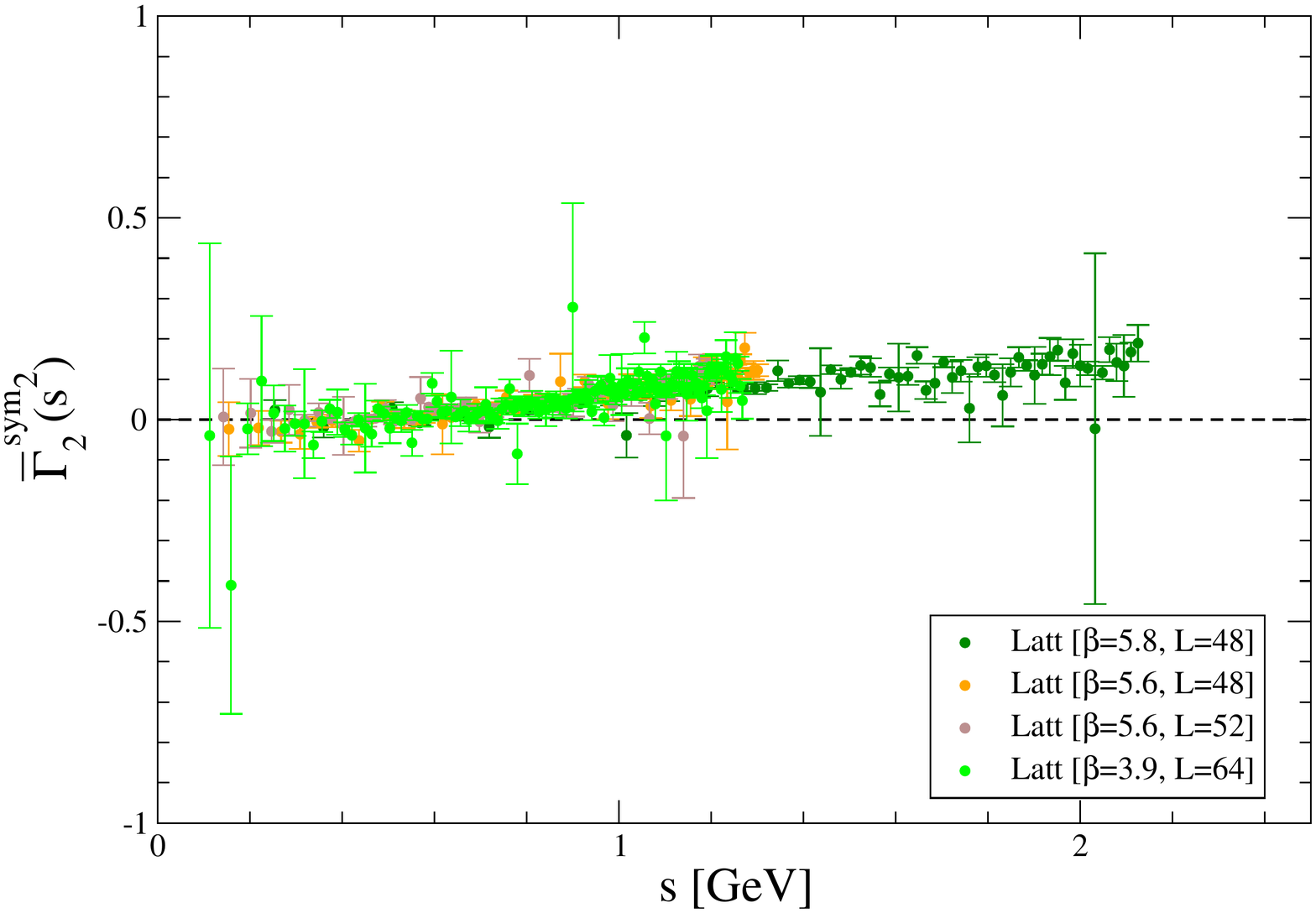} 
\vspace*{-0.55cm}
\end{tabular}
\caption{ Results for $\overline{\Gamma}_{1}^{\rm sym}(s^2)$ (upper panel) and $\overline{\Gamma}_{2}^{\rm sym}(s^2)$ (lower) obtained from three simulations with the Wilson action in a $48^4$ lattice at \mbox{$\beta$=5.8} (dark green) and 
\mbox{$\beta$=5.6} (orange), and a $52^4$ lattice at \mbox{$\beta$=5.6} (brown); and from a fourth simulation with the tree-level Symanzik action in a $64^4$ lattice at \mbox{$\beta$=3.9} (light green). Data for $\overline{\Gamma}_{1}^{\rm sym}$ covering a range of larger momenta (turquoise), obtained from several $\beta$'s and volumes and previously published in \cite{Boucaud:2002fx,Boucaud:2003xi}, have been also used to fix the subtraction point at \mbox{$\mu$=4.3 GeV}.}
\label{fig:Gamma1and2}
\end{figure}

\begin{figure}[!th]
\includegraphics[width=1.05\linewidth]{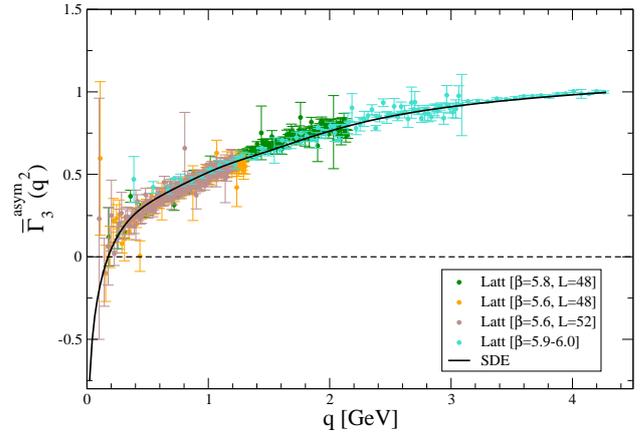} 
\vspace*{-0.55cm}
\caption{ Results for $\overline{\Gamma}_{3}^{\rm asym}(q^2)$ 
obtained from the same lattice simulations with 
the Wilson action quoted in the caption of Fig. ~\ref{fig:Gamma1and2} (same color code).  The black solid line corresponds to the SDE-based computation which, in the asymmetric limit, determines entirely the transversally projected three-gluon vertex.}
\label{fig:Gamma3}
\end{figure}

The new lattice results are shown in Figs.\,\ref{fig:Gamma1and2} and~\ref{fig:Gamma3}.
It should be stressed that the results for 
$\overline{\Gamma}_{1}^{\rm sym}$ and $\overline{\Gamma}_{3}^{\rm asym}$ are 
considerably improved with respect to  previous analyses~\cite{Athenodorou:2016oyh,Boucaud:2017obn}, 
capitalizing on a better statistical sample and the careful treatment of discretization artifacts, especially for propagators~\cite{Boucaud:2018xup}.
In addition, to the best of our knowledge, results for 
the form factor $\overline{\Gamma}_{2}^{\rm sym}$ are presented for the first time in this letter.

As a very apparent and distinctive feature, $\overline{\Gamma}_{1}^{\rm sym}$ and $\overline{\Gamma}_{3}^{\rm asym}$ clearly display the infrared suppression previously reported~\mbox{\cite{Athenodorou:2016oyh,Boucaud:2017obn,Aguilar:2019uob}}, accompanied by the characteristic logarithmic divergence near the origin. 
Instead, $\overline{\Gamma}_{2}^{\rm sym}$ appears to saturate at a small negative constant at low momenta.
As we explain below, these behaviors are well understood within the context of the SDE analysis of the previous section. 

Quite interestingly, the transverse form factors $Y_i$ of the full vertex do not contribute to the projection $\overline{\Gamma}_3^{\rm asym}(q^2)$ in the asymmetric limit, which is thus completely determined through \1eq{eq:G3XsYs} by $X_1(q^2,q^2,0)$ and $X_3(q^2,q^2,0)$.
Substituting in it the results of Fig.\,\ref{fig:X1_X3}, we obtain 
the SDE based prediction for $\overline{\Gamma}_{3}^{\rm asym}(q^2)$, given by the black continuous curve in Fig.\,\ref{fig:Gamma3};
evidently,  the coincidence with the lattice data is rather notable.
This fine agreement may be ultimately attributed to the accurate determination of $J(q^2)$ over the full range of momenta, following the
considerations outlined in section 3.  

\begin{figure}[!b]
\includegraphics[width=1.\linewidth]{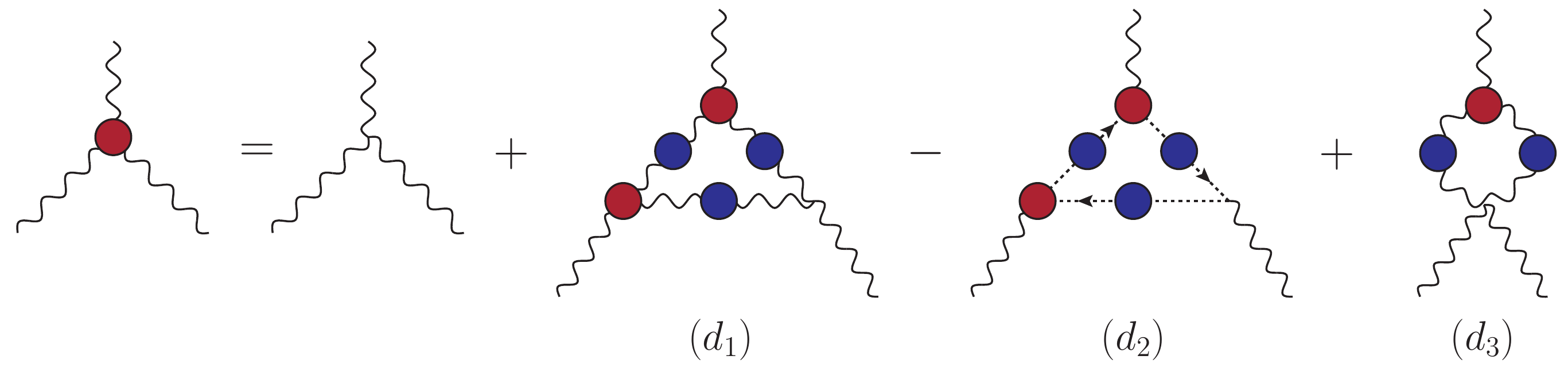}
\vspace*{-0.55cm}
\caption{ The SDE of the three-gluon vertex at the one-loop dressed level.
Blue (red) circles indicate fully dressed propagators (vertices).}
\label{fig:3g_sde2}
\end{figure}

\begin{figure}[!t]
\vspace*{-0.65cm}
\begin{tabular}{c}
\includegraphics[width=1.05\linewidth]{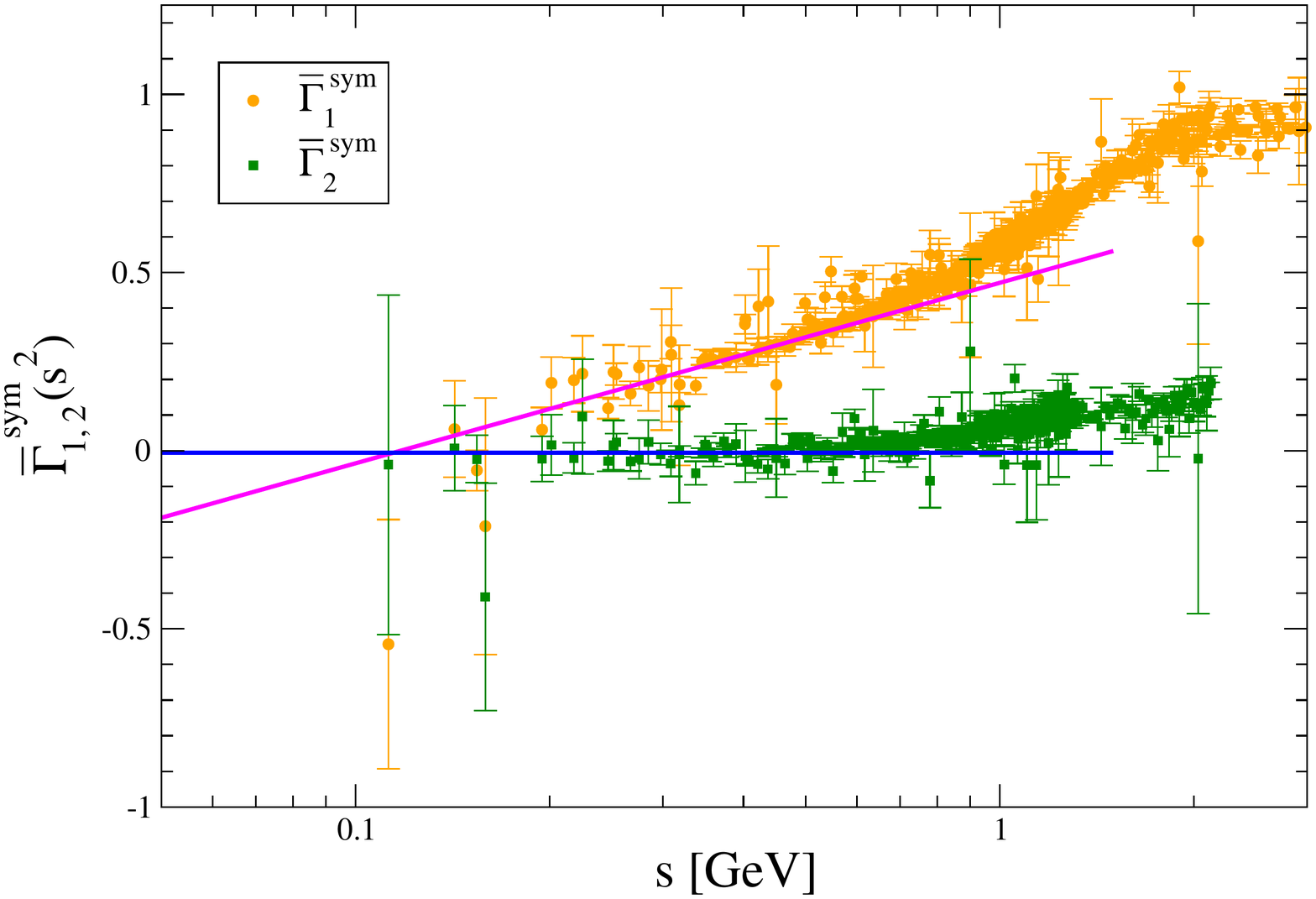} 
\vspace*{-1cm} \\
\includegraphics[width=1.05\linewidth]{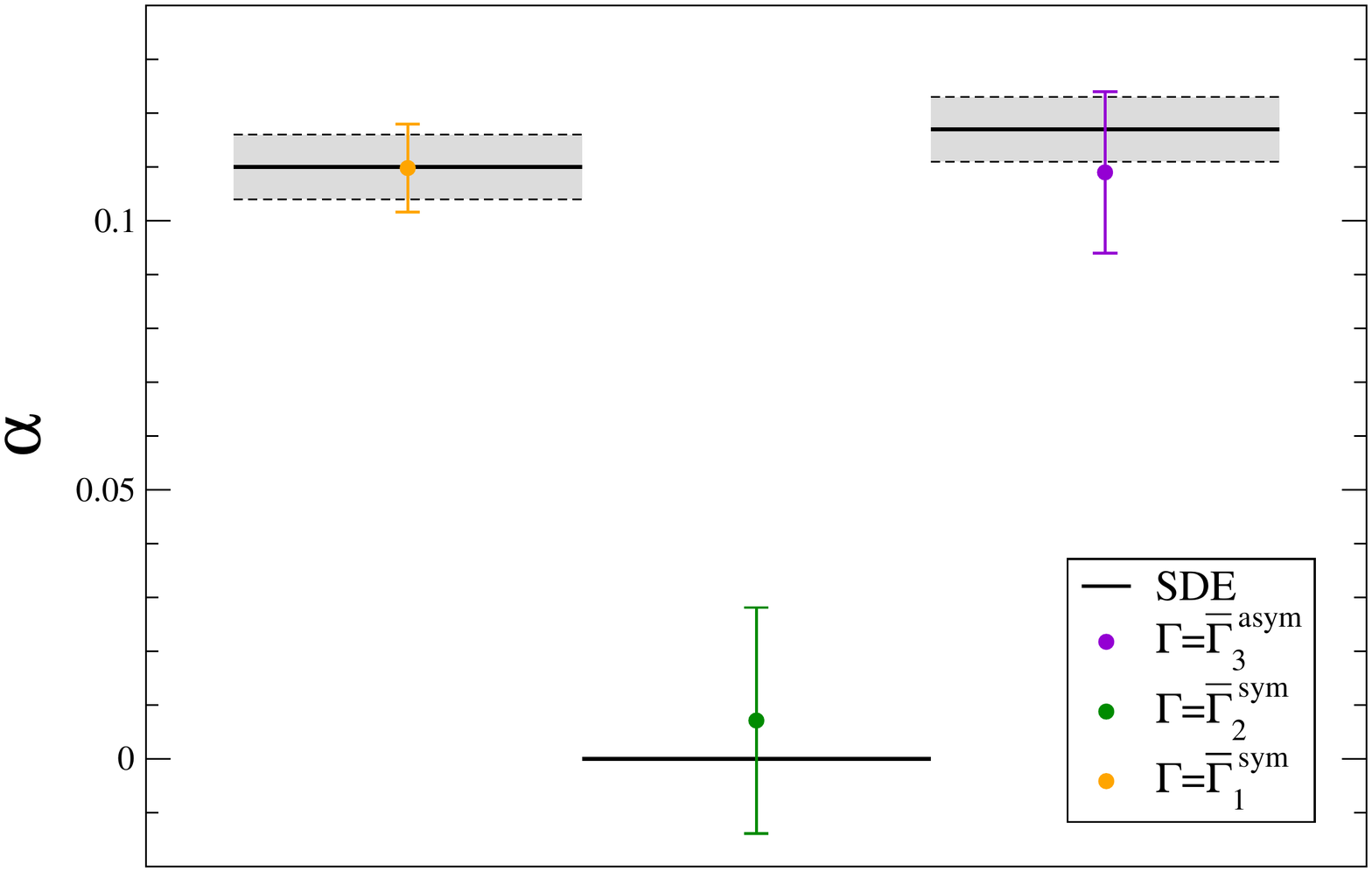} 
\vspace*{-0.55cm}
\end{tabular}
\caption{ [upper panel]\, $\overline{\Gamma}_{1}^{\rm sym} (s^2)$ (orange) and $\overline{\Gamma}_{2}^{\rm sym}(q^2)$ (green) plotted in terms of the momentum in logarithmic scale. The solid magenta and blue lines show the asymptotic infrared behavior 
  for the two form factors according, respectively, to Eqs.\,\eqref{eq:Gammapole} and\,\eqref{eq:Gamma2nopole}, the former supplemented by an intercept fitted as explained in the text.
[lower panel] The  logarithmic derivatives of $\overline{\Gamma}_{1,2}^{\rm sym} (s^2)$ and $\overline{\Gamma}_{3}^{\rm asym}(q^2)$ computed from the fit of $f(x)=\alpha\ln{x}+\beta$ [$x=q^2,s^2$] to lattice data (solid circles), with errors representing the statistical deviation from these fits, compared to their SDE estimates
  in Eqs.\,\eqref{eq:Gammapole} and\,\eqref{eq:Gamma2nopole} (solid lines). The grey bands for $\overline{\Gamma}_{1}^{\rm sym}(s^2)$ and  $\overline{\Gamma}_{3}^{\rm asym}(q^2)$ display the uncertainty obtained from the SDE value of $a$, by propagating in it a systematic error of  5 \%. }
\label{fig:Gamma1and2log}
\end{figure}

We next focus on the features of the  $\overline{\Gamma}_{i}$ in the deep infrared, contrasting the common 
behavior of $\overline{\Gamma}_{1}^{\rm sym}$ and $\overline{\Gamma}_{3}^{\rm asym}$ to that of
$\overline{\Gamma}_{2}^{\rm sym}$.
The upshot of this comparison is that, while the former quantities  
display the infrared divergence known from previous studies, the latter saturates at
a finite constant.

To that end, we turn 
to the SDE for the three-gluon vertex, shown in Fig.\,\ref{fig:3g_sde2},
and study the transverse form factors $Y_i$, which are not accessible through the STI-based
construction of the previous section. 
In particular, a detailed analysis in the symmetric limit 
reveals that,  as \mbox{$s^2\to 0$,  $Y_1(s^2) \sim c/s^4$} and \mbox{$Y_4(s^2) \sim d/s^2$},  for constants 
\mbox{$c \approx -0.07$} and \mbox{$d \approx -0.20$},  and,  consequently,    
the combinations $s^4Y_1$ and $s^2Y_4$ appearing in \1eq{eq:GsXsYs}
approach constant values at the origin.

The approximation of the $Y_i(s^2)$ through the SDE of Fig.\,\ref{fig:3g_sde2} proceeds as follows. First, the projectors that extract the scalar form factors $Y_i$ from the tensor structure of the full vertex were determined algebraically. Then, it was verified that the diagram $(d_3)$ and its permutations do not contribute to the $Y_i$ as long as the four-gluon vertex entering there is kept at tree level. The ghost-gluon and three-gluon vertices appearing in diagrams $(d_1)$ and $(d_2)$ are then approximated by retaining only form factors that possess a nonvanishing tree-level value.

Specifically, for the three-gluon vertex, we keep only the form factors $X_1$, $X_4$ and $X_7$ [see \1eq{eq:GammaLyT}], whereas the ghost-gluon vertex is approximated to $\Gamma_\mu(q,p,r) = q_\mu B_1(q,p,r)$, where $q$, $r$ and $p$, denote the momenta of the anti-ghost, ghost and gluon, respectively. Furthermore, to simplify the numerical treatment, these form factors are all considered as functions of the single momentum scale, $s^2$, and evaluated in their corresponding totally symmetric limits. Finally, for the ghost and gluon propagators we use fits to lattice data, and for the form factors $X_i(s^2)$ and $B_1(s^2)$ we use the results of Refs~\cite{Aguilar:2019jsj,Aguilar:2018csq}. 

Thus, one reaches the conclusion that the only term that furnishes logarithmically divergent contributions to the $\overline{\Gamma}_{i}$
through \2eqs{eq:GsXsYs}{eq:G3XsYs} is $X_1$, while all others provide numerical constants, {\it i.e.},
\begin{align}\label{eq:X1}
&X_1(s^2) \;  \assymto{s^2 \to 0}  \; Z_1^{\rm sym} F(s^2) J(s^2)  \; \assymto{s^2 \to 0}  \; Z_1^{\rm sym} F(0) \left[ a \ln (s^2/\mu^2)   + b  \right] \,, \nonumber\\
& s^2 X_3(s^2) \; \assymto{s^2 \to 0}\  \; - Z_1^{\rm sym} F(s^2)  s^2 J^\prime(s^2) \;  \assymto{s^2 \to 0}  \;  - Z_1^{\rm sym} F(0) \, a \,, \nonumber\\
& s^4 Y_1(s^2) \; \assymto{s^2 \to 0}\  \; c \,,
\qquad s^2 Y_4(s^2) \; \assymto{s^2 \to 0}\  \; d \,,
\end{align}

Then, from \3eqs{eq:GsXsYs}{eq:G3XsYs}{eq:X1} follows that the {\it leading} infrared contributions
of $\overline{\Gamma}_{1}^{\rm sym}(s^2)$ and $\overline{\Gamma}_{3}^{\rm asym}(q^2)$ are given by
\begin{align}\label{eq:Gammapole}
& \overline{\Gamma}_{1}^{\rm sym}(s^2) \;  \assymto{s^2 \to 0}  \; Z_1^{\rm sym} F(0)\, a \ln (s^2/\mu^2) 
 \approx 0.110(6)  \, \ln (s^2/\mu^2)\,, \\
 &\overline{\Gamma}_{3}^{\rm asym}(q^2) \;  \assymto{q^2 \to 0}  \; Z_1^{\rm asym} F(0)\, a \ln (q^2/\mu^2)
\approx 0.117(6) \,\ln (q^2/\mu^2)\,. \nonumber
\end{align}
Evidently,  both form factors diverge logarithmically to $-\infty$ at the origin,
as captured by the corresponding Figs.\,\ref{fig:Gamma1and2} and\,\ref{fig:Gamma3}, respectively.

Instead, in the same limit, $\overline{\Gamma}_{2}^{\rm sym}(s^2)$ saturates to a negative constant near zero,   
\be\label{eq:Gamma2nopole}
\overline{\Gamma}_{2}^{\rm sym}(s^2) \;  \assymto{s^2 \to 0}  \; - \frac 3 4 \left[ Z_1^{\rm sym} F(0) \,a  + \frac c 2 + \frac d 3 \right]\,
\approx  \; - 0.006(5) \,.
\ee
In obtaining the numerical values quoted above, the finite renormalization constants $Z_1^{\rm sym}$ and $Z_1^{\rm asym}$ were evaluated perturbatively at $\mu$=4.3 GeV, where they amount to 0.85 and 0.90, respectively. Moreover, 
the standard value $F(0)$=2.8\,~\cite{Aguilar:2018csq} has been employed (for the same $\mu$). 
In addition, as mentioned above, $a \approx   0.046$, while the vertex SDE yields $c \approx   -0.07$ and $d \approx  - 0.20$.
The errors have been estimated and displayed in Eqs.\,(\ref{eq:Gammapole}) and (\ref{eq:Gamma2nopole}), for illustrative purposes, through the propagation in them of an uncertainty of 5 \% in the determination of $a$, $c$ and $d$. 
Note that the numerical difference in the logarithmic slopes of $\overline{\Gamma}_{1}^{\rm sym}$ and $\overline{\Gamma}_{3}^{\rm sym}$ in Eq.\,\eqref{eq:Gammapole} is entirely due to the difference between $Z_1^{\rm sym}$ and $Z_1^{\rm asym}$.

The asymptotic behaviors of $\overline{\Gamma}_{1}^{\rm sym}(s^2)$ and $\overline{\Gamma}_{2}^{\rm sym}(s^2)$,
given in Eqs.\,\eqref{eq:Gammapole} and \eqref{eq:Gamma2nopole},
are next compared to the lattice data; the results of this comparison are shown in the upper panel of Fig.\,\ref{fig:Gamma1and2log}.
Specifically, we introduce the function $f(x) = \alpha \ln x + \beta$, which represents a straight line on
a logarithmic plot ($x=s^2,q^2$). Then, in the case of  $\overline{\Gamma}_{1}^{\rm sym}(s^2)$, 
the slope $\alpha$ is fixed at the value predicted by Eq.\,\eqref{eq:Gammapole}, 
namely  $\alpha= 0.11$, while the value of its intercept $\beta$ (not predicted by our calculation) is adjusted 
such that one gets the best fit to lattice data below $s$=0.5 GeV; the result of this procedure 
is the magenta line.
For the case of $\overline{\Gamma}_{2}^{\rm sym}(s^2)$, one simply fixes $\alpha$ and $\beta$ at their theoretical values
$\alpha=0$ and $\beta=-0.006$ (no fitting), thus obtaining the blue line.

A second comparison involves the logarithmic slopes of  
$\overline{\Gamma}_{1,2}^{\rm sym}(s^2)$ and $\overline{\Gamma}_{3}^{\rm asym}(q^2)$. 
In particularly, we now fit the lattice data below 0.5 GeV  with $f(x)$, treating both $\alpha$ and $\beta$
as free parameters, determined by a least-squares fit, including statistical errors. The resulting values
for $\alpha$, together with the associated errors,
are then compared with the theoretical predictions, as shown in the  lower panel of Fig.\,\ref{fig:Gamma1and2log}.
In all cases, the agreement is excellent, indicating a consistent picture from both SDE and lattice computations.

\section{Conclusions}

In this work we have explored crucial nonperturbative aspects of the quenched three-gluon vertex
through the combination of new lattice data obtained from large-volume
simulations and a detailed SDE-based analysis within the PT-BFM framework.

To begin with, we have acquired a clearer view of the infrared logarithmic divergences associated with the form factors 
$\overline{\Gamma}_{1}^{\rm sym}$ and $\overline{\Gamma}_{3}^{\rm asym}$ 
by  reducing considerably the statistical errors of the lattice simulation. Thus, the presence of 
these characteristic divergences, already identified in earlier studies (see {\it e.g.}, \cite{Athenodorou:2016oyh}),
is further supported by the present data.
In addition, lattice results for the form factor $\overline{\Gamma}_{2}^{\rm sym}$ are reported here for the first time,
strongly supporting its finiteness at the origin. 

The new lattice results offer an invaluable opportunity to further scrutinize key dynamical mechanisms
from new angles and perspectives. In particular, the nonperturbative features of the Landau-gauge 
two-point sector of QCD, especially the infrared finiteness of the gluon propagator and the ghost dressing function,
are instrumental for obtaining infrared divergent 
$\overline{\Gamma}_{1}^{\rm sym}$ and $\overline{\Gamma}_{3}^{\rm asym}$, and, and the same time, a finite 
$\overline{\Gamma}_{2}^{\rm sym}$. 
The observed agreement between lattice and SDE results
clearly corroborates the physical picture put forth,
and bolsters up the confidence in the predictivity of continuous functional methods in general.

\section*{Acknowledgments}
\label{sec:acknowledgments}
The work of  A.~C.~A. is supported by the CNPq grant 307854/2019-1 and the project 464898/2014-5 (INCT-FNA).
A.~C.~A. and M.~N.~F.  also acknowledge financial support from  the FAPESP projects 2017/05685-2 and 2020/12795-1, respectively.
J.~P. is supported by the  Spanish MICIU grant FPA2017-84543-P,
and the  grant  Prometeo/2019/087 of the Generalitat Valenciana. 
F.~D.~S. and J.~R.~Q.~ are supported the Spanish MICINN grant PID2019-107844-GB-C2, and regional Andalusian project P18-FR-5057.


\begin{thebibliography}{68}
\expandafter\ifx\csname natexlab\endcsname\relax\def\natexlab#1{#1}\fi
\providecommand{\url}[1]{\texttt{#1}}
\providecommand{\href}[2]{#2}
\providecommand{\path}[1]{#1}
\providecommand{\DOIprefix}{doi:}
\providecommand{\ArXivprefix}{arXiv:}
\providecommand{\URLprefix}{URL: }
\providecommand{\Pubmedprefix}{pmid:}
\providecommand{\doi}[1]{\href{http://dx.doi.org/#1}{\path{#1}}}
\providecommand{\Pubmed}[1]{\href{pmid:#1}{\path{#1}}}
\providecommand{\bibinfo}[2]{#2}
\ifx\xfnm\relax \def\xfnm[#1]{\unskip,\space#1}\fi
\bibitem[{Marciano and Pagels(1978)}]{Marciano:1977su}
\bibinfo{author}{W.~J. Marciano}, \bibinfo{author}{H.~Pagels},
  \bibinfo{journal}{Phys. Rept.} \bibinfo{volume}{36} (\bibinfo{year}{1978})
  \bibinfo{pages}{137}.
\bibitem[{Ball and Chiu(1980)}]{Ball:1980ax}
\bibinfo{author}{J.~S. Ball}, \bibinfo{author}{T.-W. Chiu},
  \bibinfo{journal}{Phys. Rev.} \bibinfo{volume}{D22} (\bibinfo{year}{1980})
  \bibinfo{pages}{2550}.
\bibitem[{Davydychev et~al.(1996)Davydychev, Osland, and
  Tarasov}]{Davydychev:1996pb}
\bibinfo{author}{A.~I. Davydychev}, \bibinfo{author}{P.~Osland},
  \bibinfo{author}{O.~V. Tarasov}, \bibinfo{journal}{Phys. Rev.}
  \bibinfo{volume}{D54} (\bibinfo{year}{1996}) \bibinfo{pages}{4087--4113}.
\bibitem[{Alkofer et~al.(2005)Alkofer, Fischer, and
  Llanes-Estrada}]{Alkofer:2004it}
\bibinfo{author}{R.~Alkofer}, \bibinfo{author}{C.~S. Fischer},
  \bibinfo{author}{F.~J. Llanes-Estrada}, \bibinfo{journal}{Phys. Lett.}
  \bibinfo{volume}{B611} (\bibinfo{year}{2005}) \bibinfo{pages}{279--288}.
\bibitem[{Cucchieri et~al.(2006)Cucchieri, Maas, and Mendes}]{Cucchieri:2006tf}
\bibinfo{author}{A.~Cucchieri}, \bibinfo{author}{A.~Maas},
  \bibinfo{author}{T.~Mendes}, \bibinfo{journal}{Phys. Rev.}
  \bibinfo{volume}{D74} (\bibinfo{year}{2006}) \bibinfo{pages}{014503}.
\bibitem[{Cucchieri et~al.(2008)Cucchieri, Maas, and Mendes}]{Cucchieri:2008qm}
\bibinfo{author}{A.~Cucchieri}, \bibinfo{author}{A.~Maas},
  \bibinfo{author}{T.~Mendes}, \bibinfo{journal}{Phys. Rev.}
  \bibinfo{volume}{D77} (\bibinfo{year}{2008}) \bibinfo{pages}{094510}.
\bibitem[{Huber et~al.(2012)Huber, Maas, and von Smekal}]{Huber:2012zj}
\bibinfo{author}{M.~Q. Huber}, \bibinfo{author}{A.~Maas},
  \bibinfo{author}{L.~von Smekal}, \bibinfo{journal}{JHEP} \bibinfo{volume}{11}
  (\bibinfo{year}{2012}) \bibinfo{pages}{035}.
\bibitem[{Pelaez et~al.(2013)Pelaez, Tissier, and Wschebor}]{Pelaez:2013cpa}
\bibinfo{author}{M.~Pelaez}, \bibinfo{author}{M.~Tissier},
  \bibinfo{author}{N.~Wschebor}, \bibinfo{journal}{Phys. Rev.}
  \bibinfo{volume}{D88} (\bibinfo{year}{2013}) \bibinfo{pages}{125003}.
\bibitem[{Aguilar et~al.(2014)Aguilar, Binosi, Iba{\~n}ez, and
  Papavassiliou}]{Aguilar:2013vaa}
\bibinfo{author}{A.~C. Aguilar}, \bibinfo{author}{D.~Binosi},
  \bibinfo{author}{D.~Iba{\~n}ez}, \bibinfo{author}{J.~Papavassiliou},
  \bibinfo{journal}{Phys. Rev.} \bibinfo{volume}{D89} (\bibinfo{year}{2014})
  \bibinfo{pages}{085008}.
\bibitem[{Blum et~al.(2014)Blum, Huber, Mitter, and von Smekal}]{Blum:2014gna}
\bibinfo{author}{A.~Blum}, \bibinfo{author}{M.~Q. Huber},
  \bibinfo{author}{M.~Mitter}, \bibinfo{author}{L.~von Smekal},
  \bibinfo{journal}{Phys. Rev.} \bibinfo{volume}{D89} (\bibinfo{year}{2014})
  \bibinfo{pages}{061703}.
\bibitem[{Eichmann et~al.(2014)Eichmann, Williams, Alkofer, and
  Vujinovic}]{Eichmann:2014xya}
\bibinfo{author}{G.~Eichmann}, \bibinfo{author}{R.~Williams},
  \bibinfo{author}{R.~Alkofer}, \bibinfo{author}{M.~Vujinovic},
  \bibinfo{journal}{Phys. Rev.} \bibinfo{volume}{D89} (\bibinfo{year}{2014})
  \bibinfo{pages}{105014}.
\bibitem[{Mitter et~al.(2015)Mitter, Pawlowski, and
  Strodthoff}]{Mitter:2014wpa}
\bibinfo{author}{M.~Mitter}, \bibinfo{author}{J.~M. Pawlowski},
  \bibinfo{author}{N.~Strodthoff}, \bibinfo{journal}{Phys. Rev.}
  \bibinfo{volume}{D91} (\bibinfo{year}{2015}) \bibinfo{pages}{054035}.
\bibitem[{Williams et~al.(2016)Williams, Fischer, and
  Heupel}]{Williams:2015cvx}
\bibinfo{author}{R.~Williams}, \bibinfo{author}{C.~S. Fischer},
  \bibinfo{author}{W.~Heupel}, \bibinfo{journal}{Phys. Rev.}
  \bibinfo{volume}{D93} (\bibinfo{year}{2016}) \bibinfo{pages}{034026}.
\bibitem[{Blum et~al.(2015)Blum, Alkofer, Huber, and Windisch}]{Blum:2015lsa}
\bibinfo{author}{A.~L. Blum}, \bibinfo{author}{R.~Alkofer},
  \bibinfo{author}{M.~Q. Huber}, \bibinfo{author}{A.~Windisch},
  \bibinfo{journal}{Acta Phys. Polon. Supp.} \bibinfo{volume}{8}
  (\bibinfo{year}{2015}) \bibinfo{pages}{321}.
\bibitem[{Cyrol et~al.(2016)Cyrol, Fister, Mitter, Pawlowski, and
  Strodthoff}]{Cyrol:2016tym}
\bibinfo{author}{A.~K. Cyrol}, \bibinfo{author}{L.~Fister},
  \bibinfo{author}{M.~Mitter}, \bibinfo{author}{J.~M. Pawlowski},
  \bibinfo{author}{N.~Strodthoff}, \bibinfo{journal}{Phys. Rev.}
  \bibinfo{volume}{D94} (\bibinfo{year}{2016}) \bibinfo{pages}{054005}.
\bibitem[{Athenodorou et~al.(2016)Athenodorou, Binosi, Boucaud, De~Soto,
  Papavassiliou, Rodriguez-Quintero, and Zafeiropoulos}]{Athenodorou:2016oyh}
\bibinfo{author}{A.~Athenodorou}, \bibinfo{author}{D.~Binosi},
  \bibinfo{author}{P.~Boucaud}, \bibinfo{author}{F.~De~Soto},
  \bibinfo{author}{J.~Papavassiliou}, \bibinfo{author}{J.~Rodriguez-Quintero},
  \bibinfo{author}{S.~Zafeiropoulos}, \bibinfo{journal}{Phys. Lett.}
  \bibinfo{volume}{B761} (\bibinfo{year}{2016}) \bibinfo{pages}{444--449}.
\bibitem[{Duarte et~al.(2016)Duarte, Oliveira, and Silva}]{Duarte:2016ieu}
\bibinfo{author}{A.~G. Duarte}, \bibinfo{author}{O.~Oliveira},
  \bibinfo{author}{P.~J. Silva}, \bibinfo{journal}{Phys. Rev.}
  \bibinfo{volume}{D94} (\bibinfo{year}{2016}) \bibinfo{pages}{074502}.
\bibitem[{Corell et~al.(2018)Corell, Cyrol, Mitter, Pawlowski, and
  Strodthoff}]{Corell:2018yil}
\bibinfo{author}{L.~Corell}, \bibinfo{author}{A.~K. Cyrol},
  \bibinfo{author}{M.~Mitter}, \bibinfo{author}{J.~M. Pawlowski},
  \bibinfo{author}{N.~Strodthoff}, \bibinfo{journal}{SciPost Phys.}
  \bibinfo{volume}{5} (\bibinfo{year}{2018}) \bibinfo{pages}{066}.
\bibitem[{Boucaud et~al.(2017)Boucaud, De~Soto, Rodríguez-Quintero, and
  Zafeiropoulos}]{Boucaud:2017obn}
\bibinfo{author}{P.~Boucaud}, \bibinfo{author}{F.~De~Soto},
  \bibinfo{author}{J.~Rodríguez-Quintero}, \bibinfo{author}{S.~Zafeiropoulos},
  \bibinfo{journal}{Phys. Rev.} \bibinfo{volume}{D95} (\bibinfo{year}{2017})
  \bibinfo{pages}{114503}.
\bibitem[{Aguilar et~al.(2019)Aguilar, Ferreira, Figueiredo, and
  Papavassiliou}]{Aguilar:2019jsj}
\bibinfo{author}{A.~C. Aguilar}, \bibinfo{author}{M.~N. Ferreira},
  \bibinfo{author}{C.~T. Figueiredo}, \bibinfo{author}{J.~Papavassiliou},
  \bibinfo{journal}{Phys. Rev.} \bibinfo{volume}{D99} (\bibinfo{year}{2019})
  \bibinfo{pages}{094010}.
\bibitem[{Aguilar et~al.(2020)Aguilar, De~Soto, Ferreira, Papavassiliou,
  Rodr\'{\i}guez-Quintero, and Zafeiropoulos}]{Aguilar:2019uob}
\bibinfo{author}{A.~C. Aguilar}, \bibinfo{author}{F.~De~Soto},
  \bibinfo{author}{M.~N. Ferreira}, \bibinfo{author}{J.~Papavassiliou},
  \bibinfo{author}{J.~Rodr\'{\i}guez-Quintero},
  \bibinfo{author}{S.~Zafeiropoulos}, \bibinfo{journal}{Eur. Phys. J. C}
  \bibinfo{volume}{80} (\bibinfo{year}{2020}) \bibinfo{pages}{154}.
\bibitem[{Aguilar et~al.(2019)Aguilar, Ferreira, Figueiredo, and
  Papavassiliou}]{Aguilar:2019kxz}
\bibinfo{author}{A.~C. Aguilar}, \bibinfo{author}{M.~N. Ferreira},
  \bibinfo{author}{C.~T. Figueiredo}, \bibinfo{author}{J.~Papavassiliou},
  \bibinfo{journal}{Phys. Rev. D} \bibinfo{volume}{100} (\bibinfo{year}{2019})
  \bibinfo{pages}{094039}.
\bibitem[{Vujinovic and Mendes(2019)}]{Vujinovic:2018nqc}
\bibinfo{author}{M.~Vujinovic}, \bibinfo{author}{T.~Mendes},
  \bibinfo{journal}{Phys. Rev.} \bibinfo{volume}{D99} (\bibinfo{year}{2019})
  \bibinfo{pages}{034501}.
\bibitem[{Cornwall(1982)}]{Cornwall:1981zr}
\bibinfo{author}{J.~M. Cornwall}, \bibinfo{journal}{Phys. Rev.}
  \bibinfo{volume}{D26} (\bibinfo{year}{1982}) \bibinfo{pages}{1453}.
\bibitem[{Bernard(1983)}]{Bernard:1982my}
\bibinfo{author}{C.~W. Bernard}, \bibinfo{journal}{Nucl. Phys.}
  \bibinfo{volume}{B219} (\bibinfo{year}{1983}) \bibinfo{pages}{341}.
\bibitem[{Donoghue(1984)}]{Donoghue:1983fy}
\bibinfo{author}{J.~F. Donoghue}, \bibinfo{journal}{Phys. Rev.}
  \bibinfo{volume}{D29} (\bibinfo{year}{1984}) \bibinfo{pages}{2559}.
\bibitem[{Wilson et~al.(1994)Wilson, Walhout, Harindranath, Zhang, Perry, and
  Glazek}]{Wilson:1994fk}
\bibinfo{author}{K.~G. Wilson}, \bibinfo{author}{T.~S. Walhout},
  \bibinfo{author}{A.~Harindranath}, \bibinfo{author}{W.-M. Zhang},
  \bibinfo{author}{R.~J. Perry}, \bibinfo{author}{S.~D. Glazek},
  \bibinfo{journal}{Phys. Rev.} \bibinfo{volume}{D49} (\bibinfo{year}{1994})
  \bibinfo{pages}{6720--6766}.
\bibitem[{Philipsen(2002)}]{Philipsen:2001ip}
\bibinfo{author}{O.~Philipsen}, \bibinfo{journal}{Nucl. Phys.}
  \bibinfo{volume}{B628} (\bibinfo{year}{2002}) \bibinfo{pages}{167--192}.
\bibitem[{Cucchieri and Mendes(2007)}]{Cucchieri:2007md}
\bibinfo{author}{A.~Cucchieri}, \bibinfo{author}{T.~Mendes},
  \bibinfo{journal}{PoS} \bibinfo{volume}{LAT2007} (\bibinfo{year}{2007})
  \bibinfo{pages}{297}.
\bibitem[{Bogolubsky et~al.(2007)Bogolubsky, Ilgenfritz, Muller-Preussker, and
  Sternbeck}]{Bogolubsky:2007ud}
\bibinfo{author}{I.~L. Bogolubsky}, \bibinfo{author}{E.~M. Ilgenfritz},
  \bibinfo{author}{M.~Muller-Preussker}, \bibinfo{author}{A.~Sternbeck},
  \bibinfo{journal}{PoS} \bibinfo{volume}{LATTICE2007} (\bibinfo{year}{2007})
  \bibinfo{pages}{290}.
\bibitem[{Bogolubsky et~al.(2009)Bogolubsky, Ilgenfritz, Muller-Preussker, and
  Sternbeck}]{Bogolubsky:2009dc}
\bibinfo{author}{I.~Bogolubsky}, \bibinfo{author}{E.~Ilgenfritz},
  \bibinfo{author}{M.~Muller-Preussker}, \bibinfo{author}{A.~Sternbeck},
  \bibinfo{journal}{Phys. Lett.} \bibinfo{volume}{B676} (\bibinfo{year}{2009})
  \bibinfo{pages}{69--73}.
\bibitem[{Oliveira and Silva(2009)}]{Oliveira:2009eh}
\bibinfo{author}{O.~Oliveira}, \bibinfo{author}{P.~Silva},
  \bibinfo{journal}{PoS} \bibinfo{volume}{LAT2009} (\bibinfo{year}{2009})
  \bibinfo{pages}{226}.
\bibitem[{Ayala et~al.(2012)Ayala, Bashir, Binosi, Cristoforetti, and
  Rodriguez-Quintero}]{Ayala:2012pb}
\bibinfo{author}{A.~Ayala}, \bibinfo{author}{A.~Bashir},
  \bibinfo{author}{D.~Binosi}, \bibinfo{author}{M.~Cristoforetti},
  \bibinfo{author}{J.~Rodriguez-Quintero}, \bibinfo{journal}{Phys. Rev.}
  \bibinfo{volume}{D86} (\bibinfo{year}{2012}) \bibinfo{pages}{074512}.
\bibitem[{Aguilar and Natale(2004)}]{Aguilar:2004sw}
\bibinfo{author}{A.~C. Aguilar}, \bibinfo{author}{A.~A. Natale},
  \bibinfo{journal}{JHEP} \bibinfo{volume}{08} (\bibinfo{year}{2004})
  \bibinfo{pages}{057}.
\bibitem[{Aguilar and Papavassiliou(2006)}]{Aguilar:2006gr}
\bibinfo{author}{A.~C. Aguilar}, \bibinfo{author}{J.~Papavassiliou},
  \bibinfo{journal}{JHEP} \bibinfo{volume}{12} (\bibinfo{year}{2006})
  \bibinfo{pages}{012}.
\bibitem[{Aguilar et~al.(2008)Aguilar, Binosi, and
  Papavassiliou}]{Aguilar:2008xm}
\bibinfo{author}{A.~C. Aguilar}, \bibinfo{author}{D.~Binosi},
  \bibinfo{author}{J.~Papavassiliou}, \bibinfo{journal}{Phys. Rev.}
  \bibinfo{volume}{D78} (\bibinfo{year}{2008}) \bibinfo{pages}{025010}.
\bibitem[{Boucaud et~al.(2008)Boucaud, Leroy, A., Micheli, P{\`{e}}ne, and
  Rodr{\'{\i}}guez-Quintero}]{Boucaud:2008ky}
\bibinfo{author}{P.~Boucaud}, \bibinfo{author}{J.~Leroy},
  \bibinfo{author}{L.~Y. A.}, \bibinfo{author}{J.~Micheli},
  \bibinfo{author}{O.~P{\`{e}}ne},
  \bibinfo{author}{J.~Rodr{\'{\i}}guez-Quintero}, \bibinfo{journal}{JHEP}
  \bibinfo{volume}{06} (\bibinfo{year}{2008}) \bibinfo{pages}{099}.
\bibitem[{Fischer et~al.(2009)Fischer, Maas, and Pawlowski}]{Fischer:2008uz}
\bibinfo{author}{C.~S. Fischer}, \bibinfo{author}{A.~Maas},
  \bibinfo{author}{J.~M. Pawlowski}, \bibinfo{journal}{Annals Phys.}
  \bibinfo{volume}{324} (\bibinfo{year}{2009}) \bibinfo{pages}{2408--2437}.
\bibitem[{Dudal et~al.(2008)Dudal, Gracey, Sorella, Vandersickel, and
  Verschelde}]{Dudal:2008sp}
\bibinfo{author}{D.~Dudal}, \bibinfo{author}{J.~A. Gracey},
  \bibinfo{author}{S.~P. Sorella}, \bibinfo{author}{N.~Vandersickel},
  \bibinfo{author}{H.~Verschelde}, \bibinfo{journal}{Phys. Rev.}
  \bibinfo{volume}{D78} (\bibinfo{year}{2008}) \bibinfo{pages}{065047}.
\bibitem[{Rodriguez-Quintero(2011)}]{RodriguezQuintero:2010wy}
\bibinfo{author}{J.~Rodriguez-Quintero}, \bibinfo{journal}{JHEP}
  \bibinfo{volume}{1101} (\bibinfo{year}{2011}) \bibinfo{pages}{105}.
\bibitem[{Tissier and Wschebor(2010)}]{Tissier:2010ts}
\bibinfo{author}{M.~Tissier}, \bibinfo{author}{N.~Wschebor},
  \bibinfo{journal}{Phys. Rev.} \bibinfo{volume}{D82} (\bibinfo{year}{2010})
  \bibinfo{pages}{101701}.
\bibitem[{Pennington and Wilson(2011)}]{Pennington:2011xs}
\bibinfo{author}{M.~Pennington}, \bibinfo{author}{D.~Wilson},
  \bibinfo{journal}{Phys. Rev.} \bibinfo{volume}{D84} (\bibinfo{year}{2011})
  \bibinfo{pages}{119901}.
\bibitem[{Cloet and Roberts(2014)}]{Cloet:2013jya}
\bibinfo{author}{I.~C. Cloet}, \bibinfo{author}{C.~D. Roberts},
  \bibinfo{journal}{Prog. Part. Nucl. Phys.} \bibinfo{volume}{77}
  (\bibinfo{year}{2014}) \bibinfo{pages}{1--69}.
\bibitem[{Fister and Pawlowski(2013)}]{Fister:2013bh}
\bibinfo{author}{L.~Fister}, \bibinfo{author}{J.~M. Pawlowski},
  \bibinfo{journal}{Phys. Rev.} \bibinfo{volume}{D88} (\bibinfo{year}{2013})
  \bibinfo{pages}{045010}.
\bibitem[{Cyrol et~al.(2015)Cyrol, Huber, and von Smekal}]{Cyrol:2014kca}
\bibinfo{author}{A.~K. Cyrol}, \bibinfo{author}{M.~Q. Huber},
  \bibinfo{author}{L.~von Smekal}, \bibinfo{journal}{Eur. Phys. J.}
  \bibinfo{volume}{C75} (\bibinfo{year}{2015}) \bibinfo{pages}{102}.
\bibitem[{Binosi et~al.(2015)Binosi, Chang, Papavassiliou, and
  Roberts}]{Binosi:2014aea}
\bibinfo{author}{D.~Binosi}, \bibinfo{author}{L.~Chang},
  \bibinfo{author}{J.~Papavassiliou}, \bibinfo{author}{C.~D. Roberts},
  \bibinfo{journal}{Phys. Lett.} \bibinfo{volume}{B742} (\bibinfo{year}{2015})
  \bibinfo{pages}{183--188}.
\bibitem[{Cyrol et~al.(2018)Cyrol, Pawlowski, Rothkopf, and
  Wink}]{Cyrol:2018xeq}
\bibinfo{author}{A.~K. Cyrol}, \bibinfo{author}{J.~M. Pawlowski},
  \bibinfo{author}{A.~Rothkopf}, \bibinfo{author}{N.~Wink},
  \bibinfo{journal}{SciPost Phys.} \bibinfo{volume}{5} (\bibinfo{year}{2018})
  \bibinfo{pages}{065}.
\bibitem[{Alkofer and von Smekal(2001)}]{Alkofer:2000wg}
\bibinfo{author}{R.~Alkofer}, \bibinfo{author}{L.~von Smekal},
  \bibinfo{journal}{Phys. Rept.} \bibinfo{volume}{353} (\bibinfo{year}{2001})
  \bibinfo{pages}{281}.
\bibitem[{Fischer(2006)}]{Fischer:2006ub}
\bibinfo{author}{C.~S. Fischer}, \bibinfo{journal}{J. Phys.}
  \bibinfo{volume}{G32} (\bibinfo{year}{2006}) \bibinfo{pages}{R253--R291}.
\bibitem[{Boucaud et~al.(2008)Boucaud, Leroy, Yaouanc, Micheli, Pene
  et~al.}]{Boucaud:2008ji}
\bibinfo{author}{P.~Boucaud}, \bibinfo{author}{J.-P. Leroy},
  \bibinfo{author}{A.~L. Yaouanc}, \bibinfo{author}{J.~Micheli},
  \bibinfo{author}{O.~Pene}, et~al., \bibinfo{journal}{JHEP}
  \bibinfo{volume}{0806} (\bibinfo{year}{2008}) \bibinfo{pages}{012}.
\bibitem[{Cornwall and Papavassiliou(1989)}]{Cornwall:1989gv}
\bibinfo{author}{J.~M. Cornwall}, \bibinfo{author}{J.~Papavassiliou},
  \bibinfo{journal}{Phys. Rev.} \bibinfo{volume}{D40} (\bibinfo{year}{1989})
  \bibinfo{pages}{3474}.
\bibitem[{Pilaftsis(1997)}]{Pilaftsis:1996fh}
\bibinfo{author}{A.~Pilaftsis}, \bibinfo{journal}{Nucl. Phys.}
  \bibinfo{volume}{B487} (\bibinfo{year}{1997}) \bibinfo{pages}{467--491}.
\bibitem[{Binosi and Papavassiliou(2009)}]{Binosi:2009qm}
\bibinfo{author}{D.~Binosi}, \bibinfo{author}{J.~Papavassiliou},
  \bibinfo{journal}{Phys. Rept.} \bibinfo{volume}{479} (\bibinfo{year}{2009})
  \bibinfo{pages}{1--152}.
\bibitem[{Abbott(1981)}]{Abbott:1980hw}
\bibinfo{author}{L.~F. Abbott}, \bibinfo{journal}{Nucl. Phys.}
  \bibinfo{volume}{B185} (\bibinfo{year}{1981}) \bibinfo{pages}{189}.
\bibitem[{Binosi and Papavassiliou(2008)}]{Binosi:2007pi}
\bibinfo{author}{D.~Binosi}, \bibinfo{author}{J.~Papavassiliou},
  \bibinfo{journal}{Phys. Rev.} \bibinfo{volume}{D77} (\bibinfo{year}{2008})
  \bibinfo{pages}{061702}.
\bibitem[{Gracey et~al.(2019)Gracey, Ki\ss{}ler, and Kreimer}]{Gracey:2019mix}
\bibinfo{author}{J.~A. Gracey}, \bibinfo{author}{H.~Ki\ss{}ler},
  \bibinfo{author}{D.~Kreimer}, \bibinfo{journal}{Phys. Rev. D}
  \bibinfo{volume}{100} (\bibinfo{year}{2019}) \bibinfo{pages}{085001}.
\bibitem[{Alles et~al.(1997)Alles, Henty, Panagopoulos, Parrinello, Pittori,
  and Richards}]{Alles:1996ka}
\bibinfo{author}{B.~Alles}, \bibinfo{author}{D.~Henty},
  \bibinfo{author}{H.~Panagopoulos}, \bibinfo{author}{C.~Parrinello},
  \bibinfo{author}{C.~Pittori}, \bibinfo{author}{D.~G. Richards},
  \bibinfo{journal}{Nucl. Phys. B} \bibinfo{volume}{502} (\bibinfo{year}{1997})
  \bibinfo{pages}{325--342}.
\bibitem[{Boucaud et~al.(1998{\natexlab{a}})Boucaud, Leroy, Micheli, Pene, and
  Roiesnel}]{Boucaud:1998bq}
\bibinfo{author}{P.~Boucaud}, \bibinfo{author}{J.~P. Leroy},
  \bibinfo{author}{J.~Micheli}, \bibinfo{author}{O.~Pene},
  \bibinfo{author}{C.~Roiesnel}, \bibinfo{journal}{JHEP} \bibinfo{volume}{10}
  (\bibinfo{year}{1998}{\natexlab{a}}) \bibinfo{pages}{017}.
\bibitem[{Boucaud et~al.(1998{\natexlab{b}})Boucaud, Leroy, Micheli, Pene, and
  Roiesnel}]{Boucaud:1998xi}
\bibinfo{author}{P.~Boucaud}, \bibinfo{author}{J.~P. Leroy},
  \bibinfo{author}{J.~Micheli}, \bibinfo{author}{O.~Pene},
  \bibinfo{author}{C.~Roiesnel}, \bibinfo{journal}{JHEP} \bibinfo{volume}{12}
  (\bibinfo{year}{1998}{\natexlab{b}}) \bibinfo{pages}{004}.
\bibitem[{Hasenfratz and Hasenfratz(1980)}]{Hasenfratz:1980kn}
\bibinfo{author}{A.~Hasenfratz}, \bibinfo{author}{P.~Hasenfratz},
  \bibinfo{journal}{Phys. Lett. B} \bibinfo{volume}{63} (\bibinfo{year}{1980})
  \bibinfo{pages}{165}.
\bibitem[{Davydychev et~al.(1998)Davydychev, Osland, and
  Tarasov}]{Davydychev:1997vh}
\bibinfo{author}{A.~I. Davydychev}, \bibinfo{author}{P.~Osland},
  \bibinfo{author}{O.~V. Tarasov}, \bibinfo{journal}{Phys. Rev.}
  \bibinfo{volume}{D58} (\bibinfo{year}{1998}) \bibinfo{pages}{036007}.
\bibitem[{Binosi et~al.(2012)Binosi, Iba\~nez, and
  Papavassiliou}]{Binosi:2012sj}
\bibinfo{author}{D.~Binosi}, \bibinfo{author}{D.~Iba\~nez},
  \bibinfo{author}{J.~Papavassiliou}, \bibinfo{journal}{Phys. Rev.}
  \bibinfo{volume}{D86} (\bibinfo{year}{2012}) \bibinfo{pages}{085033}.
\bibitem[{Aguilar et~al.(2019)Aguilar, Ferreira, Figueiredo, and
  Papavassiliou}]{Aguilar:2018csq}
\bibinfo{author}{A.~C. Aguilar}, \bibinfo{author}{M.~N. Ferreira},
  \bibinfo{author}{C.~T. Figueiredo}, \bibinfo{author}{J.~Papavassiliou},
  \bibinfo{journal}{Phys. Rev.} \bibinfo{volume}{D99} (\bibinfo{year}{2019})
  \bibinfo{pages}{034026}.
\bibitem[{Aguilar et~al.(2012)Aguilar, Ibanez, Mathieu, and
  Papavassiliou}]{Aguilar:2011xe}
\bibinfo{author}{A.~C. Aguilar}, \bibinfo{author}{D.~Ibanez},
  \bibinfo{author}{V.~Mathieu}, \bibinfo{author}{J.~Papavassiliou},
  \bibinfo{journal}{Phys. Rev.} \bibinfo{volume}{D85} (\bibinfo{year}{2012})
  \bibinfo{pages}{014018}.
\bibitem[{Aguilar et~al.(2020)Aguilar, Ferreira, and
  Papavassiliou}]{Aguilar:2020yni}
\bibinfo{author}{A.~C. Aguilar}, \bibinfo{author}{M.~N. Ferreira},
  \bibinfo{author}{J.~Papavassiliou}, \bibinfo{journal}{Eur. Phys. J. C}
  \bibinfo{volume}{80} (\bibinfo{year}{2020}) \bibinfo{pages}{887}.
\bibitem[{Boucaud et~al.(2018)Boucaud, De~Soto, Raya, Rodríguez-Quintero, and
  Zafeiropoulos}]{Boucaud:2018xup}
\bibinfo{author}{P.~Boucaud}, \bibinfo{author}{F.~De~Soto},
  \bibinfo{author}{K.~Raya}, \bibinfo{author}{J.~Rodríguez-Quintero},
  \bibinfo{author}{S.~Zafeiropoulos}, \bibinfo{journal}{Phys. Rev.}
  \bibinfo{volume}{D98} (\bibinfo{year}{2018}) \bibinfo{pages}{114515}.
\bibitem[{Boucaud et~al.(2003)Boucaud, De~Soto, Le~Yaouanc, Leroy, Micheli,
  Moutarde, Pene, and Rodriguez-Quintero}]{Boucaud:2002fx}
\bibinfo{author}{P.~Boucaud}, \bibinfo{author}{F.~De~Soto},
  \bibinfo{author}{A.~Le~Yaouanc}, \bibinfo{author}{J.~P. Leroy},
  \bibinfo{author}{J.~Micheli}, \bibinfo{author}{H.~Moutarde},
  \bibinfo{author}{O.~Pene}, \bibinfo{author}{J.~Rodriguez-Quintero},
  \bibinfo{journal}{JHEP} \bibinfo{volume}{04} (\bibinfo{year}{2003})
  \bibinfo{pages}{005}.
\bibitem[{Boucaud et~al.(2004)Boucaud, De~Soto, Le~Yaouanc, Leroy, Micheli,
  Pene, and Rodriguez-Quintero}]{Boucaud:2003xi}
\bibinfo{author}{P.~Boucaud}, \bibinfo{author}{F.~De~Soto},
  \bibinfo{author}{A.~Le~Yaouanc}, \bibinfo{author}{J.~P. Leroy},
  \bibinfo{author}{J.~Micheli}, \bibinfo{author}{O.~Pene},
  \bibinfo{author}{J.~Rodriguez-Quintero}, \bibinfo{journal}{Phys. Rev. D}
  \bibinfo{volume}{70} (\bibinfo{year}{2004}) \bibinfo{pages}{114503}.

\end{thebibliography}

\end{document}